\documentclass[aps,epsfig,prl,twocolumn,superscriptaddress,showpacs]{revtex4}
\usepackage{graphicx}% Include figure files
\usepackage{dcolumn}% Align table columns on decimal point
\usepackage{bm}% bold math
\usepackage{lineno}
\usepackage{setspace}
\def\ppbar{\ensuremath{p\overline{p}}}             %ppbar
\def\ttbar{\ensuremath{t\overline{t}}}             %ttbar
\def\pt{$p_\mathrm{T}$}                          %pT
\def\et{$E_\mathrm{T}$}                          %ET
\def\met{\mbox{${\hbox{$E$\kern-0.6em\lower-.1ex\hbox{/}}}_\mathrm{T}$}} %missing ET
\def\mex{\mbox{${\hbox{$E$\kern-0.6em\lower-.1ex\hbox{/}}}_x$}} %missing Ex
\def\mey{\mbox{${\hbox{$E$\kern-0.6em\lower-.1ex\hbox{/}}}_y$}} %missing Ey
\def\mexy{\mbox{${\hbox{$E$\kern-0.6em\lower-.1ex\hbox{/}}}_{x,y}$}} %missing Exy

\def\invfb{${\rm fb^{-1}}$}
\def\gevcc{GeV/$c^2$}                   %GeV/c^2
\def\gevc{GeV/$c$}                         %GeV/c
\def\D0{D\O}                            %D0
\Roman{table}
\begin{document}

\lefthyphenmin=2
\righthyphenmin=3
\topmargin -1cm

%EVELYN - next three - FOR CDF DRAFT ONLY
%%%\pagewiselinenumbers
%%%\doublespace
%%%\raggedbottom

%\begin{document}

%\preprint{}
\title{
%EVELYN - next three - FOR CDF DRAFT ONLY -- leave in 2-column format until submission
%%%\begin{flushright}
%%%{\small \today \hfill PRL draft v3.1 \hfill /CDF/PHYS/9649 \\}
%%%\end{flushright}
First Measurement of the $b$-jet Cross Section in Events with a $W$ Boson \\
in \ppbar\ Collisions at $\sqrt{s} =$ 1.96 TeV 
}


% \author{}
% \email[]{Your e-mail address}
% \homepage[]{Your web page}
% \thanks{}
% \altaffiliation{}
% \affiliation{}
%\input{Jan_2008_Authors_Visitors1.tex}
% EVELYN -- keep this in for PRL format drafts
\affiliation{Institute of Physics, Academia Sinica, Taipei, Taiwan 11529, Republic of China} 
\affiliation{Argonne National Laboratory, Argonne, Illinois 60439} 
\affiliation{University of Athens, 157 71 Athens, Greece} 
\affiliation{Institut de Fisica d'Altes Energies, Universitat Autonoma de Barcelona, E-08193, Bellaterra (Barcelona), Spain} 
\affiliation{Baylor University, Waco, Texas  76798} 
\affiliation{Istituto Nazionale di Fisica Nucleare Bologna, $^y$University of Bologna, I-40127 Bologna, Italy} 
\affiliation{Brandeis University, Waltham, Massachusetts 02254} 
\affiliation{University of California, Davis, Davis, California  95616} 
\affiliation{University of California, Los Angeles, Los Angeles, California  90024} 
\affiliation{University of California, San Diego, La Jolla, California  92093} 
\affiliation{University of California, Santa Barbara, Santa Barbara, California 93106} 
\affiliation{Instituto de Fisica de Cantabria, CSIC-University of Cantabria, 39005 Santander, Spain} 
\affiliation{Carnegie Mellon University, Pittsburgh, PA  15213} 
\affiliation{Enrico Fermi Institute, University of Chicago, Chicago, Illinois 60637}
\affiliation{Comenius University, 842 48 Bratislava, Slovakia; Institute of Experimental Physics, 040 01 Kosice, Slovakia} 
\affiliation{Joint Institute for Nuclear Research, RU-141980 Dubna, Russia} 
\affiliation{Duke University, Durham, North Carolina  27708} 
\affiliation{Fermi National Accelerator Laboratory, Batavia, Illinois 60510} 
\affiliation{University of Florida, Gainesville, Florida  32611} 
\affiliation{Laboratori Nazionali di Frascati, Istituto Nazionale di Fisica Nucleare, I-00044 Frascati, Italy} 
\affiliation{University of Geneva, CH-1211 Geneva 4, Switzerland} 
\affiliation{Glasgow University, Glasgow G12 8QQ, United Kingdom} 
\affiliation{Harvard University, Cambridge, Massachusetts 02138} 
\affiliation{Division of High Energy Physics, Department of Physics, University of Helsinki and Helsinki Institute of Physics, FIN-00014, Helsinki, Finland} 
\affiliation{University of Illinois, Urbana, Illinois 61801} 
\affiliation{The Johns Hopkins University, Baltimore, Maryland 21218} 
\affiliation{Institut f\"{u}r Experimentelle Kernphysik, Universit\"{a}t Karlsruhe, 76128 Karlsruhe, Germany} 
\affiliation{Center for High Energy Physics: Kyungpook National University, Daegu 702-701, Korea; Seoul National University, Seoul 151-742, Korea; Sungkyunkwan University, Suwon 440-746, Korea; Korea Institute of Science and Technology Information, Daejeon, 305-806, Korea; Chonnam National University, Gwangju,500-757, Korea; Chonbuk National University, Jeonju 561-756, Korea} 
\affiliation{Ernest Orlando Lawrence Berkeley National Laboratory, Berkeley, California 94720} 
\affiliation{University of Liverpool, Liverpool L69 7ZE, United Kingdom} 
\affiliation{University College London, London WC1E 6BT, United Kingdom} 
\affiliation{Centro de Investigaciones Energeticas Medioambientales y Tecnologicas, E-28040 Madrid, Spain} 
\affiliation{Massachusetts Institute of Technology, Cambridge, Massachusetts  02139} 
\affiliation{Institute of Particle Physics: McGill University, Montr\'{e}al, Qu\'{e}bec, Canada H3A~2T8; Simon Fraser University, Burnaby, British Columbia, Canada V5A~1S6; University of Toronto, Toronto, Ontario, Canada M5S~1A7; and TRIUMF, Vancouver, British Columbia, Canada V6T~2A3} 
\affiliation{University of Michigan, Ann Arbor, Michigan 48109} 
\affiliation{Michigan State University, East Lansing, Michigan  48824}
\affiliation{Institution for Theoretical and Experimental Physics, ITEP, Moscow 117259, Russia} 
\affiliation{University of New Mexico, Albuquerque, New Mexico 87131} 
\affiliation{Northwestern University, Evanston, Illinois  60208} 
\affiliation{The Ohio State University, Columbus, Ohio  43210} 
\affiliation{Okayama University, Okayama 700-8530, Japan} 
\affiliation{Osaka City University, Osaka 588, Japan} 
\affiliation{University of Oxford, Oxford OX1 3RH, United Kingdom} 
\affiliation{Istituto Nazionale di Fisica Nucleare, Sezione di Padova-Trento, $^z$University of Padova, I-35131 Padova, Italy} 
\affiliation{LPNHE, Universite Pierre et Marie Curie/IN2P3-CNRS, UMR7585, Paris, F-75252 France} 
\affiliation{University of Pennsylvania, Philadelphia, Pennsylvania 19104}
\affiliation{Istituto Nazionale di Fisica Nucleare Pisa, $^{aa}$University of Pisa, $^{bb}$University of Siena and $^{cc}$Scuola Normale Superiore, I-56127 Pisa, Italy} 
\affiliation{University of Pittsburgh, Pittsburgh, Pennsylvania 15260} 
\affiliation{Purdue University, West Lafayette, Indiana 47907} 
\affiliation{University of Rochester, Rochester, New York 14627} 
\affiliation{The Rockefeller University, New York, New York 10021} 
\affiliation{Istituto Nazionale di Fisica Nucleare, Sezione di Roma 1, $^{dd}$Sapienza Universit\`{a} di Roma, I-00185 Roma, Italy} 

\affiliation{Rutgers University, Piscataway, New Jersey 08855} 
\affiliation{Texas A\&M University, College Station, Texas 77843} 
\affiliation{Istituto Nazionale di Fisica Nucleare Trieste/Udine, I-34100 Trieste, $^{ee}$University of Trieste/Udine, I-33100 Udine, Italy} 
\affiliation{University of Tsukuba, Tsukuba, Ibaraki 305, Japan} 
\affiliation{Tufts University, Medford, Massachusetts 02155} 
\affiliation{Waseda University, Tokyo 169, Japan} 
\affiliation{Wayne State University, Detroit, Michigan  48201} 
\affiliation{University of Wisconsin, Madison, Wisconsin 53706} 
\affiliation{Yale University, New Haven, Connecticut 06520} 
\author{T.~Aaltonen}
\affiliation{Division of High Energy Physics, Department of Physics, University of Helsinki and Helsinki Institute of Physics, FIN-00014, Helsinki, Finland}
\author{J.~Adelman}
\affiliation{Enrico Fermi Institute, University of Chicago, Chicago, Illinois 60637}
\author{T.~Akimoto}
\affiliation{University of Tsukuba, Tsukuba, Ibaraki 305, Japan}
\author{B.~\'{A}lvarez~Gonz\'{a}lez$^t$}
\affiliation{Instituto de Fisica de Cantabria, CSIC-University of Cantabria, 39005 Santander, Spain}
\author{S.~Amerio$^z$}
\affiliation{Istituto Nazionale di Fisica Nucleare, Sezione di Padova-Trento, $^z$University of Padova, I-35131 Padova, Italy} 

\author{D.~Amidei}
\affiliation{University of Michigan, Ann Arbor, Michigan 48109}
\author{A.~Anastassov}
\affiliation{Northwestern University, Evanston, Illinois  60208}
\author{A.~Annovi}
\affiliation{Laboratori Nazionali di Frascati, Istituto Nazionale di Fisica Nucleare, I-00044 Frascati, Italy}
\author{J.~Antos}
\affiliation{Comenius University, 842 48 Bratislava, Slovakia; Institute of Experimental Physics, 040 01 Kosice, Slovakia}
\author{G.~Apollinari}
\affiliation{Fermi National Accelerator Laboratory, Batavia, Illinois 60510}
% ...

\collaboration{CDF Collaboration\footnote{With visitors from $^a$University of Massachusetts Amherst, Amherst, Massachusetts 01003,
$^b$Universiteit Antwerpen, B-2610 Antwerp, Belgium, 
$^c$University of Bristol, Bristol BS8 1TL, United Kingdom,
$^d$Chinese Academy of Sciences, Beijing 100864, China, 
$^e$Istituto Nazionale di Fisica Nucleare, Sezione di Cagliari, 09042 Monserrato (Cagliari), Italy,
$^f$University of California Irvine, Irvine, CA  92697, 
$^g$University of California Santa Cruz, Santa Cruz, CA  95064, 
$^h$Cornell University, Ithaca, NY  14853, 
$^i$University of Cyprus, Nicosia CY-1678, Cyprus, 
$^j$University College Dublin, Dublin 4, Ireland,
$^k$University of Edinburgh, Edinburgh EH9 3JZ, United Kingdom, 
$^l$University of Fukui, Fukui City, Fukui Prefecture, Japan 910-0017
$^m$Kinki University, Higashi-Osaka City, Japan 577-8502
$^n$Universidad Iberoamericana, Mexico D.F., Mexico,
$^o$University of Iowa, Iowa City, IA  52242,
$^p$Queen Mary, University of London, London, E1 4NS, England,
$^q$University of Manchester, Manchester M13 9PL, England, 
$^r$Nagasaki Institute of Applied Science, Nagasaki, Japan, 
$^s$University of Notre Dame, Notre Dame, IN 46556,
$^t$University de Oviedo, E-33007 Oviedo, Spain, 
$^u$Texas Tech University, Lubbock, TX  79609, 
$^v$IFIC(CSIC-Universitat de Valencia), 46071 Valencia, Spain,
$^w$University of Virginia, Charlottesville, VA  22904,
$^x$Bergische Universit\"at Wuppertal, 42097 Wuppertal, Germany,
$^{ff}$On leave from J.~Stefan Institute, Ljubljana, Slovenia, 
}}
\noaffiliation

%\collaboration{The CDF II Collaboration}
%\homepage{http://www-cdf.fnal.gov/}

%EVELYN FOR CDF DRAFT ONLY
%%%\maketitle

\begin{abstract}
%HMMM close to 600 now....
%char   10        20        30        40        50
%2345678901234567890123456789012345678901234567890
The cross section for jets from $b$ quarks produced with a $W$ boson
has been measured in \ppbar\ collision data from 1.9~\invfb\ of
integrated luminosity recorded by the CDF II detector at the Tevatron.
The $W+b$-jets process poses a significant background in measurements
of top quark production and prominent searches for the Higgs boson.
We measure a $b$-jet cross section of $2.74 \pm
0.27\mathrm{(stat.)}\pm 0.42\mathrm{(syst.)}$~pb in association with a
single flavor of leptonic $W$ boson decay over a limited kinematic
phase space.  This measured result cannot be accommodated in several
available theoretical predictions.
\end{abstract}

% EVELYN insert suggested PACS numbers in braces on next line -- for PRL format only
%\keywords{}

%EVELYN\maketitle need to UNCOMMENT for publication
\maketitle

% \section{}
% \subsection{}
% \subsubsection{}

The measurement of associated production of a $W$ boson and one or
more jets from $b$ quarks, herein referred to as $W+b$-jet production,
provides an important test of quantum chromodynamics (QCD).  The
understanding of this process and its description by current
theoretical calculations are important since it is the largest
background to the search for the standard model Higgs boson via $WH$
production with decay $H \rightarrow b\overline{b}$~\cite{cdfwh,d0wh}, to
measurements of top quark properties via single~\cite{cdfsingletop,d0singletop}
and pair production~\cite{cdfttbarxsec,annualreviewtop,d0ttbarxsec} with decay $t
\rightarrow W b$, and to some searches for physics beyond the standard
model~\cite{NPwb}. 
%%For comparison, the $W+b$-jet cross section is an
%%order of magnitude larger than that for single top quark
%%production~\cite{singletop_th}, and two orders of magnitude larger
%%than that for $WH$ production~\cite{whcrosssection}.
%blw start cut
% Previously, measurements have been
%made of the related processes for $Z+b$-jet
%production~\cite{beateandyPRD} and $\gamma +b$-jet
%production~\cite{gammabPRL}.
% blw end cut
%%\begin{figure}[b!]
%%\centering
%%$\begin{array}{c@{\hspace{-0.0in}}c}
%%%\multicolumn{1}{l}{\mbox{\bf (a)}} &
%%%\multicolumn{1}{l}{\mbox{\bf (b)}} \\ [-0.0cm]
%%         \scalebox{1.5}{\includegraphics[width=0.15\textwidth]{wbb.pdf}} &
%%         \scalebox{1.5}{\includegraphics[width=0.12\textwidth]{wbq_improved.pdf}} \\ 
%%%EVELYN  uncomment 2 lines above for PUBLICATION FORMAT, 2 lines below for CDF DRAFT FORMAT
%%%%%        \scalebox{1.5}{\includegraphics[width=0.3\textwidth]{wbb.pdf}} &
%%%%%        \scalebox{1.5}{\includegraphics[width=0.3\textwidth]{wbq_improved.pdf}} \\
%%\end{array}$
%%\caption{Feynman diagrams for $W$+$b$-jet production.}\label{fig:diags}
%%\end{figure}

%% This paragraph still needs some work - update references
Theoretical predictions for vector boson production with associated
$b$ jets have a large uncertainty.  
%%The Feynman diagrams for
%%contributions from $q\bar{q}^{'} \to W b\bar{b}$ and $qg \to W
%%b\bar{b} q^{'}$ are shown in Fig.~\ref{fig:diags}.  
%%For $p\bar{p}$
%%collisions at a center-of-mass energy of 1.96~TeV, the inclusive $W+b$-jets cross
%%section from a summed fixed-order QCD calculation is
%%2--3~pb~\cite{Mangano:2001xp}, assuming a $Q^2$ scale equal to the
%%$M_{W}^{2} + p_{\mathrm{T},W}^{2}$ and considering one flavor of $W\to
%%\ell \nu$ decay.  
Summed fixed-order QCD calculations for
$W$+$b\bar{b}+N$-jets production are available for up to $N$=4
additional light flavor jets and take into account $b$-quark mass
effects~\cite{Mangano:2001xp}.  
%%The $W+b$-jets cross section calculation at next-to-leading
The next-to-leading order (NLO) calculations for $W+b$-jets production
in the 1-jet and 2-jet multiplicities show an enhancement over LO up
to a factor of two for certain
diagrams~\cite{Campbell:2006cu,FebresCordero:2006sj,mcfmwbjets3}.  In
order to minimize the impact of the $W+b$-jets theoretical uncertainty
in top quark property measurements and searches for $WH$ production,
the theoretical prediction for the cross section of $W+b$-jets
production is not used in the evaluation of background estimates.
Instead, the prediction from theory for the ratio of the event yields
from $W+b$ jets and $W$ + inclusive jets, corrected to match what is
measured in data control samples, is scaled to the observed cross
section of $W$+jets in data.  The systematic uncertainty on the
$W+b$-jets yield, driven by imprecise knowledge in the fraction of
jets from $b$ production, is approximately
30-40\%~\cite{cdfwh,d0wh,cdfsingletop,d0singletop,cdfttbarxsec,d0ttbarxsec,annualreviewtop}.
These uncertainties are very large compared to the small expected
cross sections of the processes mentioned above.  We therefore wish to
directly measure the $W+b$-jet cross-section with sufficient precision
to improve those background estimations. In addition, such a
measurement will provide an important constraint on the theoretical
predictions.  Finally this measurement is a complement to other 
Tevatron measurements of vector boson plus heavy flavor jet 
production\cite{cdfzbb,d0zb,cdfwc,d0wc}.

In this Letter, we describe a measurement of the $b$-jet cross
section in events with a $W$ boson in $p\bar{p}$ collisions at
$\sqrt{s}$=1.96~TeV from a data sample corresponding to an integrated
luminosity $\mathcal{L}$=1.9~\invfb\ acquired by the Collider
Detector at Fermilab (CDF II)\cite{cdfttbarxsec}.  
We select events that are consistent with the electronic or muonic decay of a $W$ boson 
and contain one or two jets.  Among the jets in these selected events, we
seek those that originate from $b$-quark production.  Hadrons that
contain a $b$ quark have
a relatively long lifetime of $\sim$1.6~ps, and a large mass of
$\sim$5.3~\gevcc~\cite{pdg}.  We exploit the $B$ hadron's long
lifetime by examining the charged particles within each jet and
attempting to reconstruct a common origin for their trajectories that
is well-displaced from the primary
\ppbar\ interaction location. 
The distance between the primary and secondary vertices corresponds to
the trajectory through which the relativistically boosted $B$ hadron
traveled during its lifetime.  The technique is commonly known as
vertex $b$-tagging.  

The $n_{\mathrm{tag}}$ tagged jets in the selected sample are not
purely from $b$ jets.  Charm hadrons and certain light flavor hadrons
have an appreciable lifetime, and hence jets containing these hadrons
can be tagged despite not originating from $b$-quark
production.  Also, the finite resolution of the CDF tracking system
can allow for spurious displaced vertices. In order to reduce
contamination from charm and light flavor/gluon jets, the requirements
on the quality of the secondary decay vertex have been optimized for
this measurement. Further, we exploit the $B$ hadron's large mass by
examining the invariant mass of the charged particles forming the
secondary decay vertex (vertex mass, $M_\mathrm{vert}$).  Vertex mass
is correlated with the mass of the parent hadron and partially
discriminates between the possible jet flavors to yield the $b$-jet
fraction, $f^{b}$. The number of $b$ jets from other processes,
$n_{\mathrm{bkg}}^{b ~\mathrm{jets}}$, is estimated for top quark
pair, single top quark, diboson, and multijet production.  The
acceptance $\mathcal{A}_{W+b}^{b ~\mathrm{jets}}$ is defined with
respect to the restricted region of phase space defined below.  The
$b$-jet identification efficiency, $\epsilon_{\mathrm{tag}}^{b}$, and
the event trigger efficiencies, $\epsilon$, are calibrated with data.
The cross section for $b$ jets times the branching fraction for one
flavor of $W\to \ell\nu$ decay is defined as

\begin{eqnarray}\label{eqn:xsec}
\sigma_{b ~\mathrm{jets}}\times \mathcal{B} (W\rightarrow \ell \nu ) & = & \frac{n_{\mathrm{tag}} \cdot f^{b} - n_{\mathrm{bkg}}^{b ~\mathrm{jets}}}{ \sum\limits_{i=e,\mu}  (\mathcal{L} \cdot \mathcal{A}_{W+b}^{b ~\mathrm{jets}} \cdot \epsilon_{\mathrm{tag}}^{b} \cdot \epsilon)_i}
\end{eqnarray}
%%
%%\begin{eqnarray}\label{eqn:xsec}
%%\sigma_{b ~\mathrm{jets}}\times \mathcal{B} (W\rightarrow \ell \nu ) & = & \frac{n_{\mathrm{tag}} \cdot f^{b} - n_{\mathrm{bkg}}^{b ~\mathrm{jets}}}{\mathcal{L} \cdot \mathcal{A}_{W+b}^{b ~\mathrm{jets}} \cdot \epsilon_{\mathrm{tag}}^{b} \cdot \epsilon}.
%%\end{eqnarray}
\noindent where the sum is over the electron and muon channels.

%% neu: new paragraph.  I disagree with what BLW cut from here..
%%\noindent 
It is important to note that we quote our result as a jet-level cross
section in order to avoid a model-dependent correction on the number
of $b$ jets per event that would be required to convert our result into
an event-level cross section.  Further the result is defined in a
restricted region of phase space for the kinematics and multiplicity
of the outgoing particles in order to make comparisons with
theoretical predictions that minimize extrapolation outside the
experimentally accessible region.  We define this restricted region of
phase space as coincident with our analysis selection criteria, namely
to events that contain one or two hadron-level jets with \et $>20$~GeV
and $|\eta| < 2.0$, an electron or muon with \pt $>20$~GeV/$c$ and
$|\eta| < 1.1$, and a neutrino with \pt $>25$~GeV/$c$~\cite{geomnote}.
We compute the theoretical predictions with these requirements imposed
as well.

%blw start cut
%
% Note:  Citations in this section should be moved up to the first
%        reference of CDF II.
%
%The data used in this measurement come from the CDF II
%detector~\cite{fullcdfref} operating at Fermilab's Tevatron collider.
%Reconstruction of the trajectories of charged particles is achieved
%with an eight-layer silicon microstrip detector~\cite{SVXNIM,ISL,L00}
%in the pseudorapidity region $|\eta|<2$, and a 96-layer drift
%chamber~\cite{COT} in the region $|\eta|<1$.  The cylindrical tracking
%detectors are in a 1.4~T axial magnetic field that allows for momentum
%determination of charged particles.  Precision silicon tracking is
%essential for the identification of long-lived hadrons from charged
%particles that intersect at a point significantly displaced from the
%primary \ppbar\ interaction point.  Electromagnetic and hadronic
%calorimeters measure the energies of particle showers.  Outside the
%calorimeters, drift chambers and scintillation counters provide muon
%identification.  Multicell gas Cerenkov counters~\cite{clc} at small
%angles around the beam pipes determine the luminosity.
%blw end cut
%blw start cut
%Leptonic $W$ decay signatures are used since the presence of an
%electron or muon from the decay of a $W$ boson provides an efficient
%event trigger source.  
%blw end cut
The data used in this measurement come from the general purpose CDF II
detector~\cite{fullcdfref} operating at Fermilab's Tevatron collider.
Detailed descriptions of the various subdetectors new for Run II can
be found elsewhere ~\cite{SVXNIM,ISL,L00,COT,clc}.  The data are
collected with a charged lepton trigger that requires an electron
(muon) candidate with $|\eta|<1.0$ (1.1) and \et $>18$~GeV (\pt
$>18$~GeV/$c$).  The identified charged lepton and the
large missing transverse energy, \met, from the undetected
neutrino provide background suppression compared to hadronic $W$
decays.  In offline event selection we require a single reconstructed
electron (muon) with \et $> 20$~GeV (\pt $ > 20$~\gevc) that is
well-isolated from other activity in the calorimeter, and \met
$>25$~GeV.  A cone-algorithm-based jet reconstruction with cone size
$R=\sqrt{\Delta
\eta^{2}+\Delta \phi^{2}}=0.4 $ is used; jet reconstruction and
calibration are described elsewhere~\cite{jetcorr}.  We require
exactly one or two jets with \et $> 20$~GeV and $|\eta| < 2.0$.
Events consistent with cosmic rays, $Z\to\ell^{+}\ell^{-}$, photon
conversions, and multi-jet QCD production are rejected~\cite{Wjetsel}.
In data from 1.9~\invfb\ of integrated luminosity, there are 175,712
events satisfying the $W$ selection and jet requirements.

The $b$-tagging criteria have been designed for this measurement to
obtain a significantly higher purity for $b$ jets and thus reduce the
overall systematic uncertainty from the model of the vertex mass
distribution. With respect to the default CDF vertex
$b$-tagging~\cite{cdfttbarxsec}, this optimized algorithm reduces the
rate for false positives of jets from light quark flavors ($u,d,s$)
and gluons by a factor of 10 and charm by a factor of 4 at the expense
of a 50\% reduction in efficiency for $b$ jets.  To be considered for
$b$-tagging, charged particle tracks within the jet cone are required
to have \pt $>0.5$~\gevc, and impact parameter significance
$\frac{d_0}{\delta_{d_0}}>3.5$, where the impact parameter $d_0$ is
the distance of closest approach of the particle track to the location
of the primary \ppbar\ interaction in the transverse plane with
respect to the beam axis, and $\delta_{d_0}$ is its
uncertainty. Particles must also have an impact parameter less than
0.15~cm, originate from within 2~cm of the primary \ppbar\ interaction
location in the $z$ coordinate~\cite{geomnote}, and have at least a
minimum number of hits from the silicon tracking detectors.  These
requirements reduce contamination from interactions with detector
material, multiple \ppbar\ interactions, and misreconstruction,
respectively. A common decay vertex is sought among subsets of these
selected particles, and if one is found that contains three or more
particles, then the decay length in the transverse plane, $L_{2D}$, is
calculated as the projection along the jet axis of the displacement of
the secondary vertex with respect to the primary \ppbar\ interaction
location.  The vertex is required to have decay length significance
$\frac{L_{2D}}{\delta_{L_{2D}}}>7.5$, and pseudo-c$\tau \equiv
\frac{L_{2D}\cdot M_{\mathrm{vert}}\cdot c}{p_\mathrm{T,
vert}}<1.0$~cm, where the invariant mass $M_{\mathrm{vert}}$ and
transverse momentum $p_{\mathrm{T, vert}}$ of the vertex are
calculated from the constituent particles.  Note that the mass of each
particle is set to the charged pion mass.  Any vertices consistent
with $K^0_S$ and $\Lambda$ decay, and nuclear interactions in the
detector material are rejected.  The sign of the vertex tag is
determined by the position of the vertex with respect to the jet
direction; those on the same (opposite) hemisphere as the jet
direction are called positively (negatively) tagged.

Among the events satisfying our event selection, $n_\mathrm{tag} =$
943 jets are found to be positively tagged. 
%%There are 18 events with two tagged jets.  
The flavor composition of the positively tagged sample is determined
through a maximum likelihood fit of the distribution of the vertex mass 
in the data.  
%%As bottom hadrons are generally more massive
%%than charm hadrons, which in turn are generally more massive than
%%purely light flavor hadrons, it follows that vertices in $b$ jets are
%%more massive than those in charm jets and in light flavor jets.
Simulated distributions for $b$ and charm jets are formed from
standard model processes that are major contributors to the selected
event sample.  Sources of $b$ jets include $W$+$b$ jets, which is
simulated by the Monte Carlo event generator {\sc alpgen} version
2.1.0~\cite{alpgen} with {\sc cteq5l} parton densities~\cite{cteq} and
{\sc pythia} version 6.325 for hadronization~\cite{pythia}; \ttbar\
and diboson production are simulated with {\sc pythia} version 6.216,
and single top quark production by {\sc madevent} version
4.2.11~\cite{madevent}.  The yields of tagged jets from these
processes are determined from the simulated samples scaled to the
latest theoretical cross sections~\cite{ttbar_th,singletop_th,diboson_th} 
with the event selection requirements applied.

\begin{figure}[ht]
\vspace{-0.4cm}
\includegraphics[angle=0,width=0.48\textwidth]{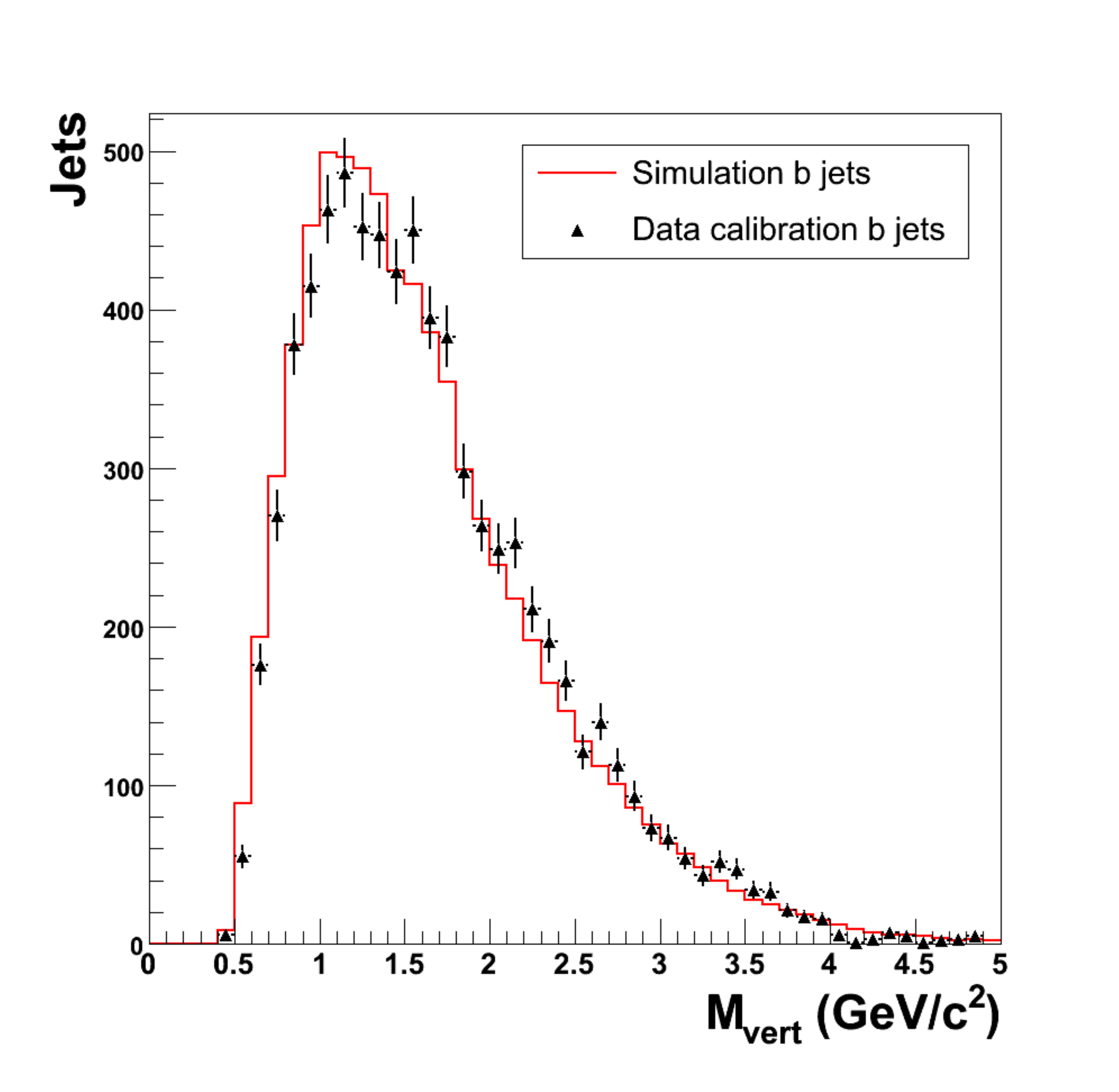}
%EVELYN  uncomment line above for PUBLICATION FORMAT, line below for CDF DRAFT FORMAT
%%%\includegraphics[angle=0,width=0.9\textwidth]{comp2_bcalibdataVmc_UT_b_mvert_single_improved8_pub.pdf}
\vspace{-1.0cm}
\caption{\small Comparison of the vertex mass distributions of tagged $b$ jets from
a simulated sample and from data.}\label{fig:bcalib}
\end{figure}

We check the simulation model of $b$ jets against an independent data
sample from double-tagged dijet events collected with a single \pt
$>9$~\gevc\ muon trigger.  One jet is required to contain the muon,
presumably from semileptonic $B$ hadron decay. The other tagged jet in
these events is a sample whose $b$-jet purity is estimated to be above
99\%.  This sample is used to validate the model of the $b$-jet vertex mass;
the agreement between simulation and data is shown
in Fig.~\ref{fig:bcalib}.
%%We varied several parameters of the
%%simulation by their uncertainties, including fragmentation and decay
%%branching fractions, and the fraction of jets containing two bottom
%%hadrons.  As no single aspect of the simulation of the $b$ jets could
%%account for the difference with respect to data, w
We use the difference between simulation and data to estimate the
systematic uncertainty from the $b$-jet model.

Vertex tags of jets from charm hadrons are primarily due to
$W$+$c$-jets production, which is simulated with {\sc alpgen}.
Positive vertex tags of light flavor jets are modeled with a
simulation of inclusive jet production from {\sc pythia}. We use
negatively tagged jets in the data as an alternative model for light
flavor.  This second light flavor model is used in the vertex mass fit
to assess the impact of light flavor model choice on the result.

\begin{figure}[t]
\vspace{-0.5cm}
\includegraphics[width=0.48\textwidth]{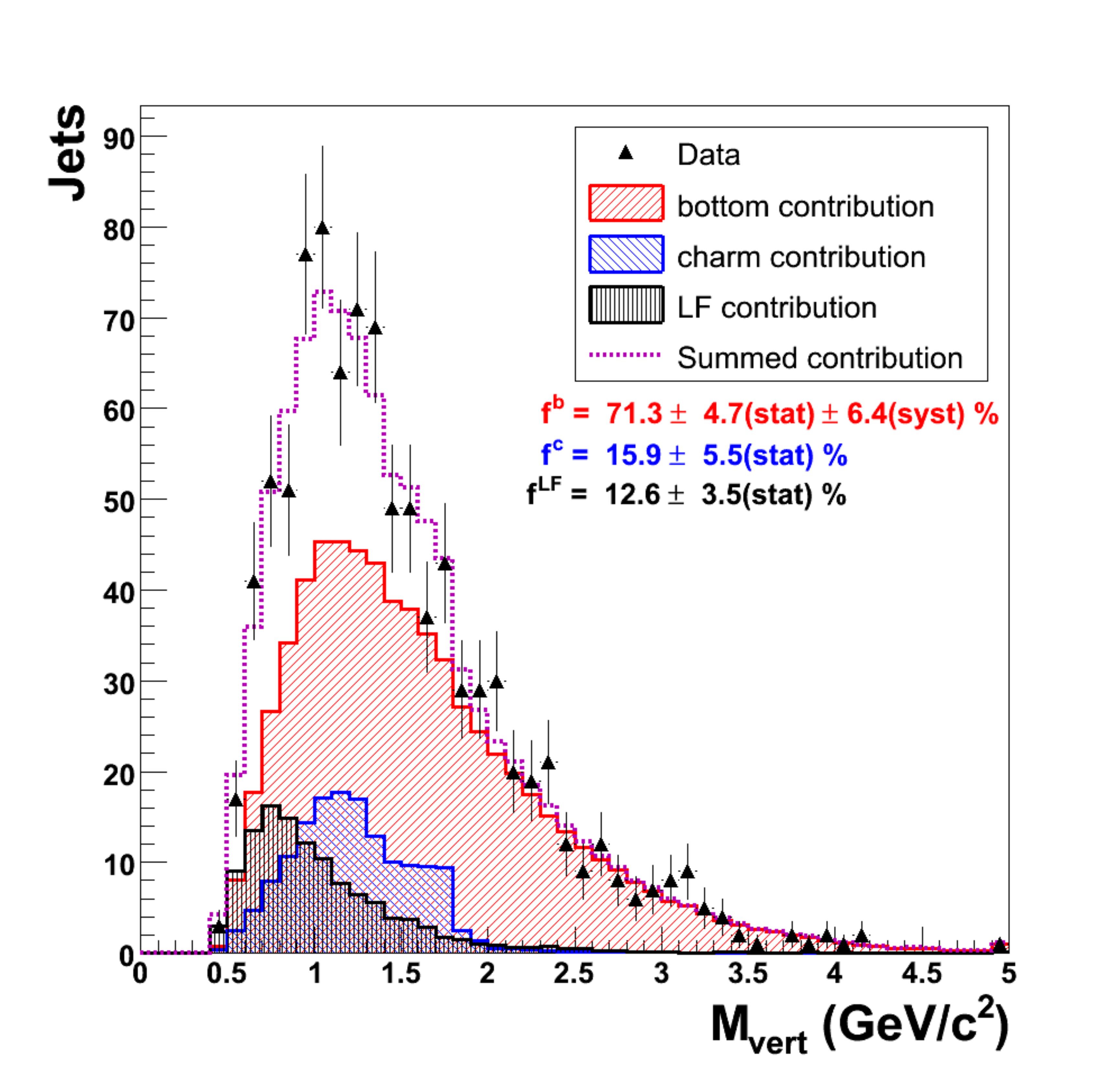}
%EVELYN  uncomment line above for PUBLICATION FORMAT, line below for CDF DRAFT FORMAT
%%%\includegraphics[width=0.9\textwidth]{data_can_allhighptemu_1900_final_improved8_pub.pdf}
\vspace{-1cm}
\caption{\small Maximum likelihood fit of the vertex mass for tagged jets in the selected data sample.
%The feature near 1.8~GeV/$c^{2}$, evident from the fitted charm contribution, is due to
%full participation in the found vertex of the charged particle tracks
%in $D^{+} \rightarrow K^{+}\pi^{-}\pi^{+}$ decays and other similar
%charmed hadron decay processes.
}\label{fig:datamvertfit}
\end{figure}

The maximum likelihood fit of the vertex mass data distribution, shown
in Fig.~\ref{fig:datamvertfit}, is used to extract two parameters: the
fraction of jets from bottom hadrons $f^b$, and the fraction of jets
from charmed hadrons $f^c$, where the fraction of jets from light
flavors is $f^\mathrm{LF} \equiv 1 - f^b - f^c$.  The best fit is
$f^{b}= 0.71 \pm 0.05$(stat.) corresponding to $670 \pm 44$(stat.) 
tagged jets from bottom hadrons.  From simulated experiments with
flavor compositions similar to the data, we confirmed that our vertex
mass fit procedure returns results consistent with the assumed
background content.  These simulated experiments indicated that the
systematic uncertainties on the model of the $b$, charm, and light
flavor vertex mass distributions manifest themselves as relative
systematic uncertainties of 0.08, 0.01 and 0.03, respectively, on the
fitted $b$-jet fraction.
%%The 8\% relative
%%uncertainty is due to the residual difference in vertex mass shape for
%%$b$ jets in data and simulation as evidenced by Fig.~\ref{fig:bcalib}.
%%We looked into several possible sources that could account for this
%%difference between data and simulation by the variation within quoted
%%uncertainties of fragmentation and decay branching fractions, and the
%%fraction of jets containing two $B$ hadrons.  No single aspect of the
%%simulated $b$ jets could account for the entire mismatch with respect
%%to data.  (now in caption) The feature near 1.8~GeV/$c^{2}$, evident from the fitted
%%charm contribution, is due to full participation in the found vertex
%%of the charged particle tracks in $D^{+} \rightarrow
%%K^{+}\pi^{-}\pi^{+}$ decays and other similar charmed hadron decay
%%processes.

%%
%% Comment: the relative uncertainties of 0.08, 0.01, and 0.03 are on the
%% the final result, xsec x BR.  Should we make the distinction?
%%
This yield of $b$ jets includes our signal but also contains a
contribution from other processes with jets from $b$-quark production.
We use simulated samples and the theoretical predictions for
production rates of $t\bar{t}$~\cite{ttbar_th}, single top
quark~\cite{singletop_th}, and diboson processes ($WZ$, $WW$, and
$ZZ$)~\cite{diboson_th} in order to estimate a contribution of $152
\pm 21$ $b$ jets from these processes.  This includes a small
contribution of 7.3 $\pm$ 0.8 jets from $W$+$b$-jets production with
$W\to\tau\nu$, which is treated as a background. Sources of systematic
uncertainty in the background yield of tagged $b$ jets include the
uncertainty in the $b$-jet tagging efficiency in the data (a relative
6\% uncertainty on all tagged $b$-jet yields), the uncertainty on the
top quark and diboson predicted cross sections (a relative uncertainty
of 10\% on \ttbar\ and diboson, and 30\% on single top yields, which
translate to an overall 2\% uncertainty on
$\sigma_{b ~\mathrm{jets}}\times \mathcal{B}$) and the uncertainty in
the accumulated CDF luminosity (a relative 6\% on all yields).

%The contribution from $W+b$-jets production but with the decay
%$W\to\tau\nu$ is treated as a background and modeled with ALPGEN.
%Similarly $Z+b$ jets, with $Z \to \ell \ell $ is also treated as a
%background...modeled with ALPGEN...these are small.

We estimate a contribution of $25 \pm 8$ $b$-jets from QCD multijet
production, where mismeasured jets pass the lepton identification
requirements and result in sufficient \met.  As this background is
difficult to model with simulation, a complementary data sample was
collected with the same high \pt\ electron trigger, but where the
electron candidate failed at least two of the identification
criteria~\cite{messinacooperpaper}.  This provides both a model of the
\met\ distribution, which is used to estimate the rate of QCD multijet
background above our selection~\cite{messinacooperpaper}, and a vertex
mass distribution, which is used to determine the fraction of tagged
jets from bottom hadrons.  The model for tagged jets from multijet
production is statistics limited; we recover statistics by relaxing
the \met\ requirement and perform the vertex mass distribution fit for
\met\ $>$ 15 and 20 GeV as well as the default \met\ $>$ 25 GeV and
use all three results to determine the fitted $b$ fraction from
multijet production. The uncertainties on the QCD multijet tagged
$b$-jet background come from the modeling of the \met\ distribution
for the overall multijet normalization (a relative 30\% uncertainty,
which translates to a 1\% uncertainty on
$\sigma_{b ~\mathrm{jets}}\times \mathcal{B}$), and the spread in the
fitted $b$ fraction from vertex mass distribution fits from the different
\met\ thresholds (a relative 25\% uncertainty, which translates to a
1\% uncertainty on $\sigma_{b ~\mathrm{jets}}\times
\mathcal{B}$).  

After subtracting the background of
$n_{\mathrm{bkg}}^{b~\mathrm{jets}} = 177 \pm 22$, we have a yield of
$493 \pm 48$(stat.) tagged $b$ jets from $W$+$b$ production.  We
define the acceptance, $\mathcal{A}_{W+b}^{b ~\mathrm{jets}}$, of our
selection with respect to a restricted region of kinematic phase
space, as defined earlier. The phase space restrictions are applied
to the outgoing leptonic $W$ daughters and jets in the simulated
$W$+$b$ production {\sc alpgen} events.  Hadron-level jets are defined
by {\sc SpartyJet}~\cite{spartyjet} as a collection of simulated final
state particles that have been clustered using the same cone algorithm
as in the jet reconstruction.  A hadron-level jet is said to be
$b$-matched if it has $\Delta R
\def\sqrt{\Delta \phi^2 + \Delta \eta^2} < 0.4$ with respect to a $b$
quark in the simulated event. The matching considers $b$-quark
candidates after showering but before hadronization. The denominator
of the acceptance is the number of $b$-matched hadron-level jets in
simulated $W+b$-jet events that pass the phase space requirements as
given earlier; the numerator is the number of $b$-matched
reconstructed jets in simulated $W+b$-jet events that pass the phase
space requirements and, in addition, the event selection described
above through the jet multiplicity requirement.  The weighted average
acceptance over the electron and muon channels is found to be
0.68$\pm$0.03, where the sources that dominate the systematic
uncertainty are the jet energy calibration (3\%), the factorization
and renormalization scale (3\%), and the dependence of event
kinematics on the parton distribution functions (2\%).

For clarity, we separate the $b$-tag efficiency and several
data-based corrections from the acceptance. The $b$-tag efficiency is
the ratio of the number of $b$-tagged reconstructed $b$-matched jets
to the number of reconstructed $b$-matched jets in the simulated
$W+b$-jet events that have passed the event selection and phase space
requirements: $\epsilon_{~\mathrm{tag}}^{b,\mathrm{sim}} = 0.177\pm
0.001$(stat.).  This value needs to be corrected by a factor of $0.88
\pm 0.01\textrm{(stat.)} \pm 0.05\textrm{(syst.)}$, which quantifies the
discrepancy in tag efficiency between simulation and
data~\cite{neubtag}.  The corrected $b$-tag efficiency is then
$\epsilon_{\mathrm{tag}}^{b} = 0.156\pm 0.009$.  The final correction
factor $\epsilon$ is the average over all triggers of the product of
the following three terms determined from data: the fraction of events
that happen in the luminous region well-contained by the CDF detector,
with primary \ppbar\ interaction within 60~cm of the center of the
detector along the beam line, 0.963 $\pm$ 0.003; the efficiency of the
trigger, 0.943 $\pm$ 0.004; and the correction factor for charged
lepton identification efficiency, 0.969 $\pm$ 0.004.

Having obtained all of the information needed as input to
Eq.~\ref{eqn:xsec}, we measure the $b$-jet cross section to be
$\sigma_{b ~\mathrm{jets}}\times \mathcal{B} (W\rightarrow \ell \nu )
= 2.74 \pm 0.27\textrm{(stat.)} \pm 0.42\textrm{(syst.)}$~pb 
with a $W$ boson decaying to a single leptonic flavor
within the restricted kinematic phase space defined
earlier.  The overall relative uncertainty on the measurement is 18\%.
This uncertainty is dominated by the uncertainty in the $b$-jet
vertex mass model (a relative 8\% on
$\sigma_{b ~\mathrm{jets}}\times\mathcal{B}$), the tag efficiency
(6\%), and the luminosity (6\%). The results in the electron and 
muon channels were examined independently as a cross check and 
are consistent.  

%blw start cut
%We
Finally, we
%blw end cut 
have determined the theoretical prediction of
$\sigma_{b ~\mathrm{jets}}\times \mathcal{B}$, using our kinematic
definition above, at leading order from {\sc pythia} and at
summed fixed-order from {\sc alpgen}.  The {\sc pythia} prediction is
1.10~pb and the {\sc alpgen} prediction is 0.78~pb, assuming a $Q^2$
scale of $M_{W}^{2} + p_{\mathrm{T},W}^{2}$; these predictions are
factors of 2.5-3.5 lower than our result.  These are important
comparisons given the wide use of these programs in the generation of
simulated physics events at the Tevatron and LHC experiments.  A NLO
calculation of $\sigma_{b ~\mathrm{jets}}\times \mathcal{B}$ has
recently been completed~\cite{campbellprov}; their prediction of $1.22
\pm 0.14\textrm{(syst.)}$~pb is also low with respect to the measured value.  Further
study is underway to examine the differential cross section as a
function of jet kinematics and compare to LO, summed fixed-order and
NLO predictions.
%blw start cut
%In conclusion, we have measured the cross section for $b$ jets in
%events with a $W$ boson. Jets from $b$-quark production are identified by the
%presence of a decay vertex significantly displaced from the primary
%interaction point. The $b$-tagging requirements employed here were
%designed specifically for this analysis.  We measure the b-jet cross
%section to be 2.74 $\pm$ 0.27(stat.) $\pm$ 0.42(syst.) pb for
%one flavor of leptonic $W$ boson decay and within the
%restricted kinematic phase space of the analysis.  Available
%fixed-order theoretical calculations predict that this process has a cross
%section 2-3 times lower than observed in our data, while a NLO 
%calculation is in good agreement with the observed result.
%blw end cut

We thank the Fermilab staff and the technical staffs of the
participating institutions for their vital contributions. This work
was supported by the U.S. Department of Energy and National Science
Foundation; the Italian Istituto Nazionale di Fisica Nucleare; the
Ministry of Education, Culture, Sports, Science and Technology of
Japan; the Natural Sciences and Engineering Research Council of
Canada; the National Science Council of the Republic of China; the
Swiss National Science Foundation; the A.P. Sloan Foundation; the
Bundesministerium f\"ur Bildung und Forschung, Germany; the Korean
Science and Engineering Foundation and the Korean Research Foundation;
the Science and Technology Facilities Council and the Royal Society,
UK; the Institut National de Physique Nucleaire et Physique des
Particules/CNRS; the Russian Foundation for Basic Research; the
Ministerio de Educaci\'{o}n y Ciencia and Programa Consolider-Ingenio
2010, Spain; the Slovak R\&D Agency; and the Academy of Finland.

\bibliography{wbjets_prl}

\begin{thebibliography}{37}
\bibitem[{geo()}]{geomnote}
\bibinfo{note}{{We use a cylindrical coordinate system in which the $z$ axis is
  along the proton beam direction and $\theta$\ is the polar angle.
  Pseudorapidity is $\eta = -\ln\tan(\theta/2)$, while transverse momentum is
  $p_T = |p| \sin \theta$, and transverse energy is $E_T = E \sin \theta$.
  Missing transverse energy, \met, is defined as the magnitude of $-\sum_i
  E_T^i \hat{n}_i$, where $\hat{n}_i$\ is the unit vector in the azimuthal
  plane that points from the beam line to the $i^{\mathrm th}$ calorimeter
  tower.}}
\end{thebibliography}

\end{document}